\documentclass[english]{article}
\usepackage[T1]{fontenc}
\usepackage{amssymb}
\usepackage{graphicx}

\makeatletter
\usepackage{a4}
\input{epsfig.sty}
\def\fnum@table{\tablename~{\bf\thetable}}
\def\fnum@figure{\figurename~{\bf\thefigure}}
\def\tablename{\footnotesize{\bf Table}}
\def\figurename{\footnotesize{\bf Figure}}

\def\be{\begin{equation}}
\def\ee{\end{equation}}

\makeatother

\usepackage{babel}
\begin{document}

\title{\textbf{Constraining high energy interaction mechanisms by 
studying forward hadron production at the LHC}}

\author{S.\ Ostapchenko$^{1,2}$, M.\ Bleicher$^{1,3}$, T.\ Pierog$^{4}$
and K.\ Werner$^{5}$\\
$^1$\textit{\small Frankfurt Institute for Advanced Studies, 
 60438 Frankfurt am Main, Germany}\\
$^2$\textit{\small D.V. Skobeltsyn Institute of Nuclear Physics,
Moscow State University, 119992 Moscow, Russia }\\
$^3$\textit{\small Institute for Theoretical Physics, Goethe-Universit\"at,
 60438 Frankfurt am Main, Germany}\\
$^4$\textit{\small Karlsruhe Institute of Technology, Institut f\"ur Kernphysik, Postfach 3640,
76021 Karlsruhe, Germany}\\
$^5$\textit{\small SUBATECH, University of Nantes--IN2P3/CNRS--EMN,  4 rue Alfred Kastler,
44307 Nantes Cedex 3, France}
}

\maketitle
\begin{center}
\textbf{Abstract}
\par\end{center}

We demonstrate that underlying assumptions concerning the structure of
constituent parton Fock states in hadrons make a strong impact on the
predictions of hadronic interaction models for forward hadron spectra
and for long-range correlations between central and forward hadron
production. Our analysis shows that combined studies of proton-proton
collisions at the Large Hadron Collider
 by central and forward-looking detectors have a rich 
potential for discriminating between the main model approaches.

\section{Introduction\label{intro.sec}}
\label{sec:intro}
The modeling of high energy hadronic interactions is of considerable importance
for experimental studies at the Large Hadron Collider (LHC) and, especially,
 in 
astroparticle physics. Working on event
 generators for hadronic and nuclear collisions, one
generally aims at describing the largest possible part of the corresponding
phase space using perturbative methods. Nevertheless, one can not restrain
from  using phenomenological  approaches when dealing with essentially
nonperturbative soft physics which unavoidably enters such developments.
This applies, in particular, to the treatment of constituent parton Fock states
in hadrons.

\begin{figure}[t]
\centering
\includegraphics[height=4.5cm,width=0.3\textwidth,clip]{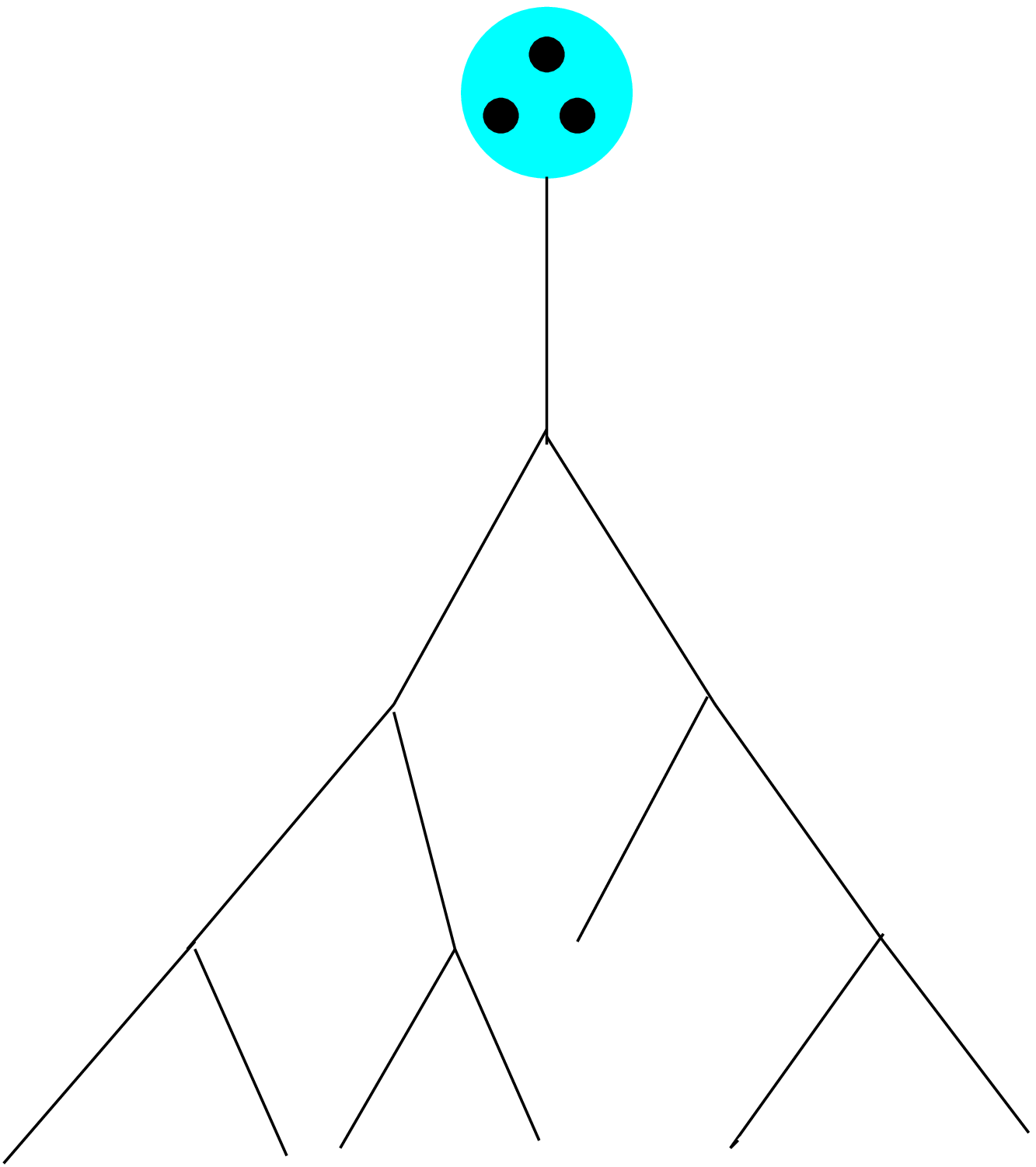}\hspace{2cm}
\includegraphics[height=4.5cm,width=0.4\textwidth,clip]{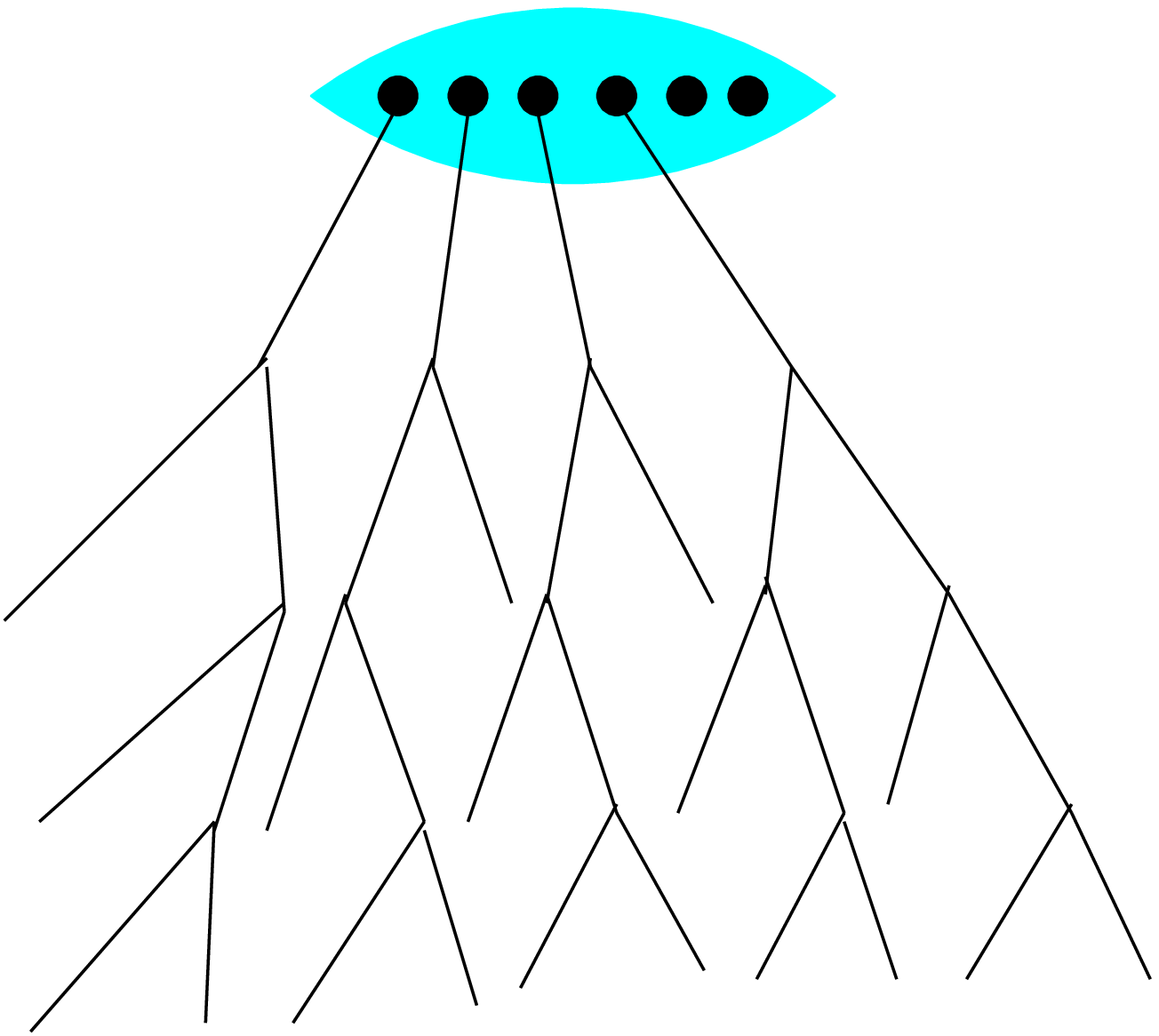}
\caption{Schematic view on the initial part of the parton cascade in the proton.
Left: the cascade starts from the same universal parton Fock state;
new partons participating in multiple scattering processes emerge from the
cascade development, being characterized by $\propto 1/x$  distributions
for the light-cone momentum fraction. Right: the proton is represented by
a superposition of Fock states consisting of different numbers of large $x$
constituent partons; the more abundant multiple scattering the larger
Fock states get involved in the process.}
\label{fig:fock}       
\end{figure}

Considering, e.g., a proton-proton  collision in the center of mass (c.m.) 
frame, we generally deal with multiple binary  interactions of partons from
  two parton clouds formed prior to the collision. We are interested here
  in the momentum distribution of these multiparton states, which is related
  to the evolution of parton cascades in hadrons. The picture behind the
corresponding treatment for the majority of   current generators of hadronic
collisions is shown schematically on the left-hand side (lhs) of
 Fig.\ \ref{fig:fock}. %
At large Feynman $x$, one usually starts from the same universal  
parton Fock state.  Additional
partons  (sea quarks and gluons)
 giving rise to new branches of the parton cascade, which take
part in the multiple scattering processes, result from the evolution of the
parton density corresponding to this initial 
state and are typically distributed as
$\propto 1/x$ in the very high energy limit. For example, the DIPSY generator
\cite{avs05} contains an explicit 
treatment of this kind, developed in the dipole framework.
Such a picture reflects itself in the hadron production pattern predicted
by the models: multiple scattering mostly affects central particle production,
while having a weak influence on forward hadron spectra. Indeed, the latter
are formed by the hadronization of partons emerging from the
initial part of the underlying parton cascade, which starts from the same
initial conditions and covers a short rapidity interval, 
being thus weakly dependent on further development of the cascade.

Alternatively, one may consider a proton to be represented by a superposition
of a number of Fock states containing different numbers of large $x$ 
constituent partons, as shown schematically on the right-hand side  of
 Fig.\ \ref{fig:fock}.
Further cascading of these partons ``dresses''  them with  low $x$ parton
clouds. As the overall parton multiplicity in the central rapidity region
is roughly proportional to the
number of initial constituent partons, stronger multiple scattering is
 typically associated with larger Fock states. Thus, there is a strong
long-range correlation between central and forward particle production; higher
 multiplicity in the central rapidity region reflects stronger multiple parton
 scattering. In turn, this implies that larger numbers of constituent partons
 are   involved  in the process, which has a strong impact on forward hadron
 spectra. This approach is typically used in models developed within the
 Reggeon field theory (RFT) framework \cite{gri68}, 
 like EPOS \cite{wer06,pie15} or QGSJET-II \cite{ost06,ost11}.
 It is noteworthy that a similar picture arises when considering the
 incident proton to be at rest and using e.g.\ the color glass condensate
 framework: In the very high energy limit, there is a dense
 cloud of partons originating from the parton cascade in the target proton;
 these partons pass through the projectile proton and undergo multiple 
 scattering off its valence constituents \cite{dre05}.

To quantify the discussion, let us consider  
  the (quasi-)eikonal RFT approach  \cite{kai82},  where the proton is
    represented by a superposition of Fock states consisting of 
    different numbers $n$ of 
constituent partons: $|p\rangle =\sum_n \sqrt{C_n}\,|n\rangle$, $C_n$ being the 
respective partial weights. The behavior
of the corresponding parton light-cone momentum fraction distributions
 $F^{(n)}(x_1,...,x_n)$
is described in the low $x$ limit by secondary Reggeon asymptotics\footnote{
We restrict our discussion to typical hadronic collisions that give rise to
 the bulk of secondary particle production. Hence, we neglect contributions 
 of  parton Fock states containing heavy quarks \cite{bro80}, which may
 generally be important for more dedicated observables (see Ref.\ \cite{bro15}
 for a recent review).}
\cite{kai87}:
\begin{equation}
\left. F^{(n)}(x_1,...,x_n)\right|_{x_i\rightarrow 0}
 \propto x_i^{-\alpha_{\mathbb R}(0)},\;\;\alpha_{\mathbb R}(0)\simeq 0.5.
\end{equation}
Thus, we deal with Fock states formed by predominantly large $x$ constituent
partons. The interaction is mediated by multi-Pomeron exchanges; the partial
contribution of an exchange by $m$ Pomerons rises with c.m.\ energy squared $s$
like
$\propto s^{m(\alpha_{\mathbb P}(0)-1)}$, where $\alpha_{\mathbb P}(0)>1$ is
the Pomeron intercept. Hence, the higher the energy the more abundant 
the multiple
scattering and Fock states with increasing numbers of partons coming
into play. Due to the so-called Abramovskii-Gribov-Kancheli (AGK) cancellations
\cite{agk},  multi-Pomeron configurations do not contribute to
the inclusive hadron spectra  in the central rapidity region 
and the latter rise  as a power of energy like 
\begin{equation}
\left.\frac{d\sigma_{pp}^h(s,y)}{dy}\right|_{y\rightarrow 0}\propto 
s^{\alpha_{\mathbb P}(0)-1}.\label{agk-middle}
\end{equation}
 Moreover, taking into account the energy-momentum sharing between 
 constituent partons at the amplitude level, the AGK cancellations apply
to the whole kinematic space \cite{hla01} and the complete hadron spectra are
characterized by the powerlike behavior:\footnote{In Eq.\ (\ref{eq:spec-agk}),
we do not include the contribution to secondary hadron production from the
hadronization of proton ``remnants'' formed by   spectator 
partons \cite{dre01}.}
\begin{equation}
\frac{E\,d^3\sigma_{pp}^h(s,p)}{dp^3}\propto 
s^{\alpha_{\mathbb P}(0)-1}.
 \label{eq:spec-agk}
\end{equation}
In turn, inclusive hadron spectra for proton-nucleus collisions are related
to the ones for $pp$ interactions as
\begin{equation}
\frac{E\,d^3\sigma_{pA}^h(s,p)}{dp^3}=
A\,\frac{E\,d^3\sigma_{pp}^h(s,p)}{dp^3}\,.
 \label{eq:spec-agk-A}
\end{equation}
Such a picture gives rise to the strongest possible correlation between
central and forward particle production: both originate from the hadronization
of Pomeron ``strings'' stretched between the projectile and target (large $x$)
constituent partons \cite{dre01}.

As discussed in Ref.\ \cite{dre01}, Eqs.\ (\ref{eq:spec-agk}) and (\ref{eq:spec-agk-A}) 
should fail at sufficiently high energies
as the total light-cone momentum of produced hadrons would
exceed the one of the incident proton in the limits of large $s$ or large $A$.
In the high energy limit, one has to take into consideration the contributions
of so-called enhanced diagrams that describe Pomeron-Pomeron interactions
\cite{kan73}. This considerably suppresses the forward hadron spectra
 compared to Eqs.\ (\ref{eq:spec-agk}) and (\ref{eq:spec-agk-A}) and weakens the 
 correlation between   central and forward particle production
 \cite{agk,dre01}. The corresponding treatment of the QGSJET-II model is based
 on an all-order resummation of the respective graphs \cite{ost06a} while
 the EPOS model employs an effective treatment of lowest order diagrams
 \cite{wer06}.

It is noteworthy that the alternative approach corresponding to the schematic
picture in the lhs  of Fig.\ \ref{fig:fock} can be recovered here if one
restricts oneself to the contributions of the enhanced diagrams only.
To be more precise, one has to consider only those graphs where  a single
Pomeron is coupled to the projectile proton, respectively, to the target
proton or  to a target nucleon in case of $pA$ collision. In the $pp$ case,
this leads to the set of the so-called Pomeron loop graphs. For 
proton-nucleus collisions, also 
 the ``fan'' diagrams developing in the target direction
 have to be considered.
 This would lead precisely to the picture  in the lhs  of Fig.\ \ref{fig:fock}. 
  While for central density of produced particles, Eq.\ (\ref{agk-middle}) will
 remain approximately valid, forward hadron spectra will scale 
 proportionally to the inelastic cross section: 
 $\left. d\sigma_{pp}^h/dx_{\rm F}\right|_{x_{\rm F}\rightarrow 1}
 \propto \sigma_{pp}^{\rm inel}(s)$, $x_{\rm F}$ being the Feynman $x$
 variable. In other words, in the fragmentation region
 Feynman $x$ scaling would hold with a good accuracy:
 \begin{equation}
\frac{1}{\sigma_{pp}^{\rm inel}(s)}\:  \left. \frac{d\sigma_{pp}^h(s,x_{\rm F})}
{dx_{\rm F}}\right|_{x_{\rm F}\rightarrow 1}\simeq f(x_{\rm F}).
 \label{eq:scaling}
\end{equation}

In this paper, we   analyze the differences between Monte Carlo generators
for high energy hadronic collisions, which employ the two above-discussed
approaches to the treatment of constituent parton Fock states in hadrons.
In Section \ref{sec:model-diff}, we compare the energy 
  dependence
 of forward hadron spectra in $pp$
 collisions, predicted
by the EPOS-LHC \cite{pie15} and QGSJET-II-04 \cite{ost11} models, to the 
respective results of the alternative treatment corresponding to 
the picture  in the lhs  of Fig.\ \ref{fig:fock}, as implemented in
 the PYTHIA~6 (Perugia tune 350) \cite{pythia,perugia}  and SIBYLL~2.3
  \cite{rie15} models. 
In Section \ref{sec:tests}, we illustrate how the two basic approaches can
be discriminated by combined measurements of hadron production at the LHC by
central and forward-looking detectors. Finally, we conclude in  
 Section \ref{sec:conclusions}.
  
\section{Energy  dependence of forward hadron spectra and of the
``inelasticity''}
\label{sec:model-diff}
An accurate model description of forward hadron production is of utmost
importance for various experimental activities in the collider 
and astroparticle
physics fields. It is relevant, for example, for studies of the inelastic
diffraction at the LHC (see, e.g., the discussion in Ref.\ \cite{ost11a}) or
to various astrophysical studies with charged cosmic rays and gamma rays,
including indirect searches for dark matter, as discussed, e.g.,
in Refs.\ \cite{kam05,kac12,mos15}. 
But of most crucial importance is the modeling of forward production for
the interpretation of experimental data on ultrahigh energy cosmic rays
 which are studied with the extensive air shower (EAS) techniques --
measuring the properties of nuclear-electromagnetic cascades induced by  
primary cosmic ray particles (protons or nuclei) in the atmosphere. Indeed,
the energy dependence of forward hadron spectra impacts strongly the relation
between the properties of the primary particles and the calculated
EAS characteristics,
in particular,  for  the so-called EAS maximum position $X_{\max}$
 -- the depth in the
atmosphere where the maximal number of ionizing particles is observed
\cite{ulr11,eng11}.  As discussed in Ref.\ \cite{ulr11}, the calculated $X_{\max}$ is
especially sensitive to model predictions for the so-called inelasticity 
 -- the relative energy loss of leading nucleons in proton-air collisions.%
\begin{figure}[t]
\centering
\includegraphics[height=10cm,width=0.9\textwidth,clip]{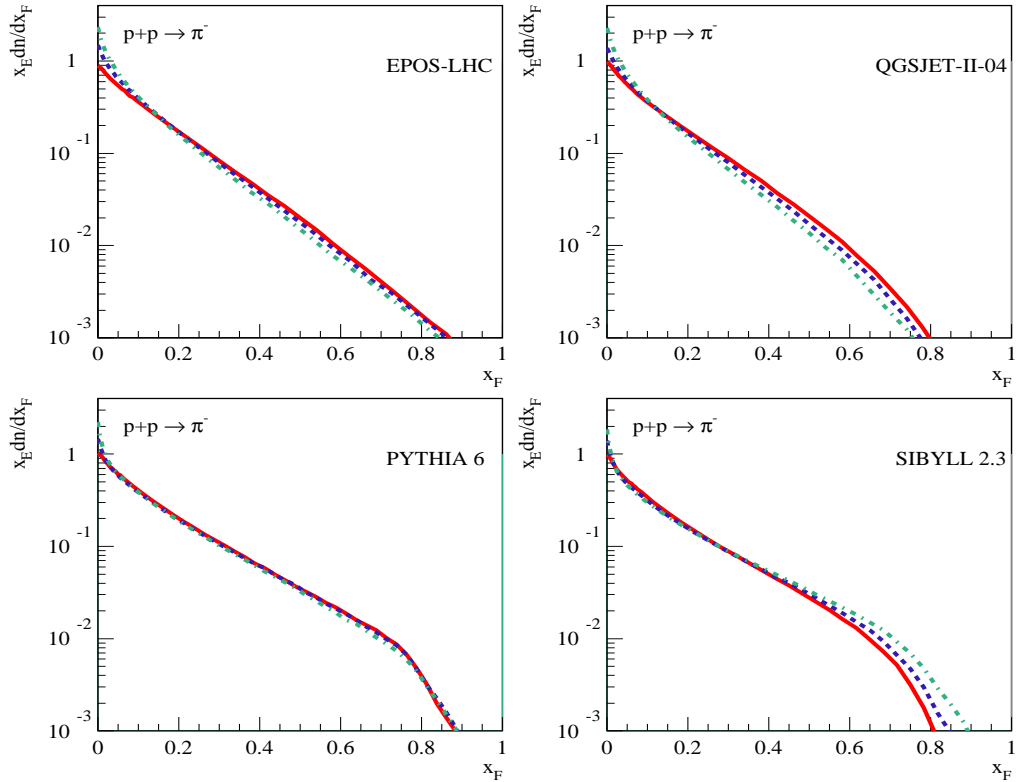}
\caption{Feynman $x$ spectra of negative pions  in $pp$ collisions at 
$\sqrt{s}=10^2$ (solid), $10^3$ (dashed), and $10^4$ (dash-dotted) GeV, 
as calculated using EPOS-LHC (top left),   QGSJET-II-04 (top right),
 PYTHIA~6 (bottom left), and  SIBYLL~2.3 (bottom right). }
\label{fig:spec-e}       
\end{figure}%

\begin{figure}[t]
\centering
\includegraphics[height=10cm,width=0.9\textwidth,clip]{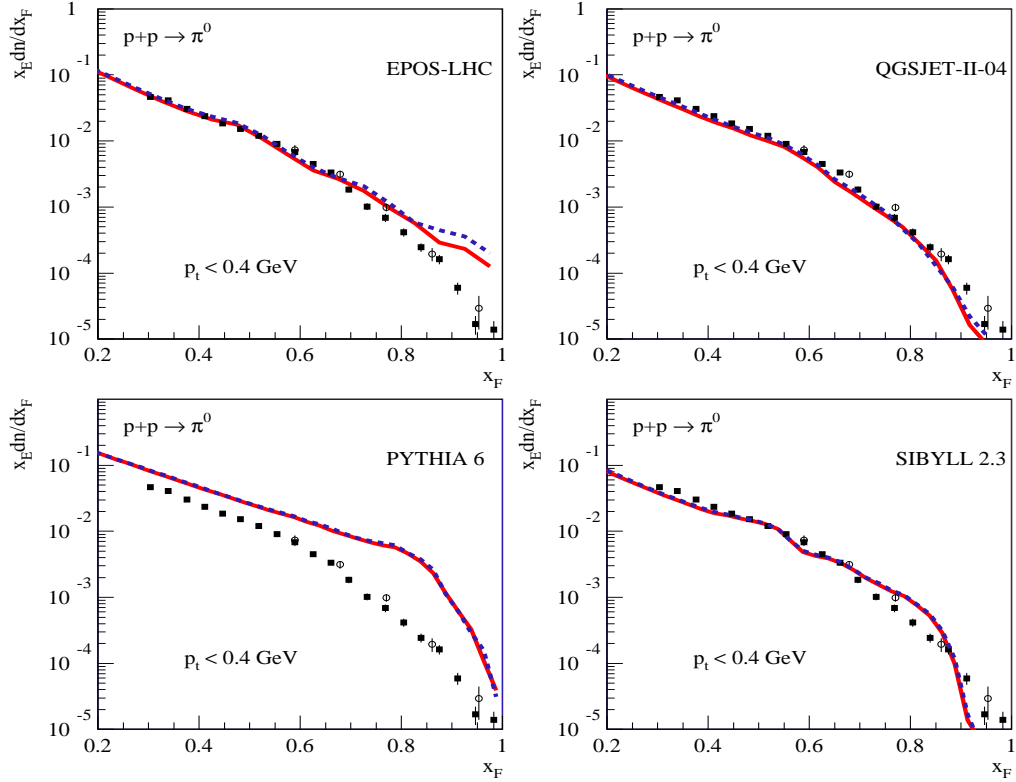}
\caption{Feynman $x$ spectra of neutral pions (for transverse momentum
 $p_{\rm t}<0.4$ GeV) in $pp$ collisions at 
$\sqrt{s}=7$ (solid)  and $2.76$ (dashed) TeV, 
as calculated using EPOS-LHC (top left),   QGSJET-II-04 (top right),
 PYTHIA~6 (bottom left), and  SIBYLL~2.3 (bottom right), compared to 
 LHCf data  \cite{lhcf-pi0} ($\sqrt{s}=7$ TeV -- filled squares,
$\sqrt{s}=2.76$ TeV --  open circles).}
\label{fig:pi0-lhcf}       
\end{figure}

The differences in the predicted energy dependence of forward hadron spectra
are illustrated in Fig.\ \ref{fig:spec-e}.
There we plot   Feynman $x$ spectra of negative
  pions $x_E\,dn_{pp}^{\pi^-}/dx_{\rm F}$ ($x_E=2E/\sqrt{s}$)
  for $pp$ collisions at $\sqrt{s}=10^2$, $10^3$, and $10^4$ GeV,
as calculated with EPOS-LHC, QGSJET-II-04, PYTHIA~6, and SIBYLL~2.3. 
While all the models predict a similar energy rise of the pion yield
in the central region   ($x_{\rm F}\simeq 0$), their results for the
forward spectra differ considerably.  For PYTHIA~6,
 the Feynman $x$ scaling of Eq.\  (\ref{eq:scaling}) holds 
to a  very good accuracy: 
 the  spectral shape is practically independent of $\sqrt{s}$
 for  $x_{\rm F}\gtrsim 0.01 $.
SIBYLL~2.3 shows a   similar  behavior up to large  $x_{\rm F}$  values;
for  $x_{\rm F}\gtrsim 0.5$, it predicts a hardening of the very 
forward  pion spectra, which is related to a specific treatment of 
 the hadronization of the proton ``remnant'' state in that model.
In contrast, for  EPOS-LHC and
 QGSJET-II-04 the predicted energy dependence is rather different.
 For   $x_{\rm F}\lesssim 0.1$, the spectra rise with   $\sqrt{s}$,
though with increasing $x_{\rm F}$ this rise is strongly  suppressed by 
absorptive effects, compared to the AGK-like 
dependence of Eq.\ (\ref{eq:spec-agk}). The suppression 
is especially strong in the very forward direction where we observe a
significant ``softening'' of the predicted spectra. 
Comparing in Fig.\ \ref{fig:pi0-lhcf} the model predictions for forward 
$\pi^0$ spectra at $\sqrt{s}=2.76$ and 7 TeV  with the respective measurements
  of the LHCf experiment   \cite{lhcf-pi0},
  we observe a  generally better agreement with the data
  for EPOS-LHC and QGSJET-II-04.
 However, the relatively small
difference between the two collision energies does not allow one to make
definite conclusions concerning the degree of the scaling violation.

\begin{figure}[t]
\centering
\includegraphics[height=5.5cm,width=0.5\textwidth,clip]{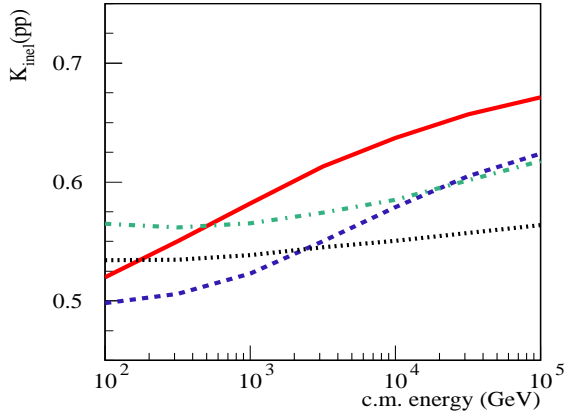}
\caption{Energy dependence of the inelasticity of leading nucleons
 in $pp$ collisions, as calculated using EPOS-LHC (solid),
 QGSJET-II-04 (dashed), PYTHIA~6 (dash-dotted), and SIBYLL~2.3 (dotted).}
\label{fig:inel}       
\end{figure}

 In Fig.\ \ref{fig:inel}, %
we show the energy dependence of the  inelasticity 
$K^{\rm inel}_{pp}$  of leading nucleons in $pp$ collisions,
 predicted by the different models. The purely AGK-like behavior corresponding
to Eq.\ (\ref{eq:spec-agk}) would give rise to a powerlike increase
of the inelasticity: $K^{\rm inel}_{pp}(s)\propto s^{\alpha_{\mathbb P}(0)-1}$.
In reality, important contributions to forward nucleon spectra come from
the hadronization of hadronic ``remnant'' states formed by spectator partons
and from the inelastic diffraction, which should modify the $s$-dependence as
 \begin{equation}
K^{\rm inel}_{pp}(s)\simeq K_0+{\rm const}\times s^{\alpha_{\mathbb
P}(0)-1}\,.\label{eq:kinel-s}
 \end{equation}
 In the very high energy limit, such a rise should be necessarily damped by
 absorptive corrections, to prevent the violation of the energy conservation
  \cite{dre01}.
In  Fig.~\ref{fig:inel}, we observe indeed a substantial increase of the
inelasticity predicted by    EPOS-LHC and  QGSJET-II-04. For the latter model,
at $\sqrt{s}\simeq 10^2 - 10^4$ GeV $K^{\rm inel}_{pp}$ manifests the 
energy dependence described by Eq.\ (\ref{eq:kinel-s}), while for   EPOS-LHC
the stronger absorptive corrections damp it to just a logarithmic energy
rise in this energy range.
  In turn, the mechanism 
corresponding   to the schematic
picture in the lhs  of Fig.~\ref{fig:fock} should result in a rather weak energy
dependence of the inelasticity,  which we observe indeed in  Fig.~\ref{fig:inel}
for the PYTHIA~6 and  SIBYLL~2.3 models.

\section{How to discriminate between the interaction mechanisms?}
\label{sec:tests}
\begin{figure}[t]
\centering
\includegraphics[height=5.5cm,width=0.9\textwidth,clip]{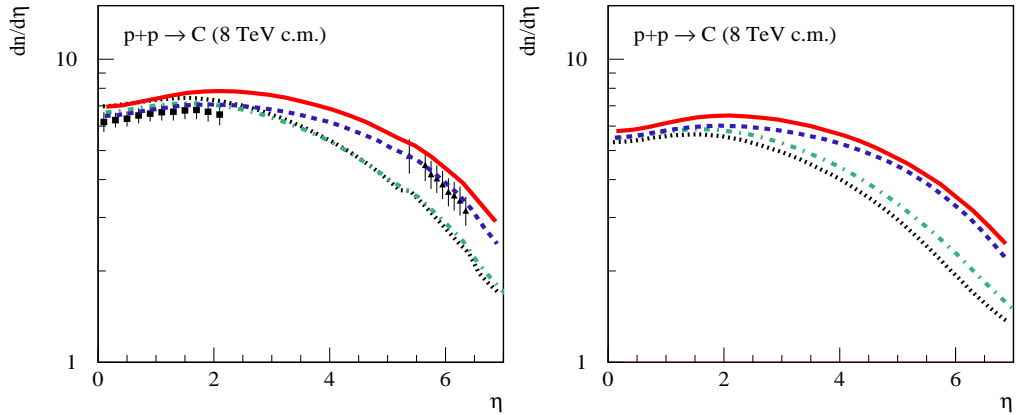}
\caption{Pseudorapidity   density of produced charged hadrons
 in $pp$ collisions at $\sqrt{s}=8$ TeV, as calculated using EPOS-LHC (solid),
 QGSJET-II-04 (dashed), PYTHIA~6 (dash-dotted), and SIBYLL~2.3 (dotted) for two
 different   event selections: at least one charged hadron produced
both  at $-6.5<\eta<-5.3$ and at $5.3<\eta<6.5$ (left panel) or 
  at least one charged hadron of $p_{\rm t}>0.1$ GeV, 
  produced at $|\eta|<2.5$ (right panel).   The data of the CMS and TOTEM 
  experiments are shown  by filled squares and filled triangles respectively.}
\label{fig:cms-tot}       
\end{figure}

Recent combined measurements by the CMS and TOTEM experiments
 of the pseudorapidity $\eta$ density    $dn^{\rm ch}_{pp}/d\eta$ 
 of produced charged hadrons in $pp$ collisions \cite{cms-tot} 
 proved to be quite sensitive to the discussed 
 interaction mechanisms. This is illustrated in Fig.\ \ref{fig:cms-tot} (left), %
where we compare the calculated $dn^{\rm ch}_{pp}/d\eta$  for the 
different models  to the CMS and TOTEM data, for 
the nondiffractive (ND) event selection adopted in the experimental
analysis:  at least one charged hadron produced both  at $-6.5<\eta<-5.3$ 
and at $5.3<\eta<6.5$. In the EPOS-LHC and  QGSJET-II-04 models, the predicted
 $dn^{\rm ch}_{pp}/d\eta$  is characterized by a relatively flat
  $\eta$-dependence, extending to large $\eta$ values, in agreement with
  the experimental observations.
   In contrast, for the other two models  the $\eta$-density
 of produced hadrons quickly falls down in the forward direction, reflecting
 the quick decrease of the number of constituent partons when parton momentum
 fraction increases.
 
 It is noteworthy that the adopted experimental selection complicates
 somewhat the model comparison -- as the fraction of ND events which satisfy
 the trigger depends noticeably on the predicted  absolute values of
 $dn^{\rm ch}_{pp}/d\eta$ at large $\eta$. Indeed, the probability for producing
 a rapidity gap of the rather small size $\Delta \eta \simeq 1$ by fluctuations
 in ND events is rather high \cite{bjo92,aad12} and varies strongly from 
 model to model (see Ref.\ \cite{kmr11} for a detailed quantitative study).
 Thus, a cleaner comparison may be
  provided by a central detector trigger,
 e.g.\ requesting at least one charged hadron of $p_{\rm t}>0.1$ GeV to
  be detected at $|\eta|<2.5$. The corresponding results plotted  in 
 Fig.~\ref{fig:cms-tot} (right) show a good agreement between all the models
 in the central $\eta$ region, which reflects the model calibration to the
 data of Run 1 of the LHC. On the other hand, the models diverge considerably
 in the forward $\eta$ region, which reflects the above-discussed differences
 concerning the assumed structure of constituent parton Fock states:
 The results of EPOS-LHC and  QGSJET-II-04 practically coincide with each
 other and are both substantially flatter compared to  $dn^{\rm ch}_{pp}/d\eta$ 
 predicted by the other two models.

\begin{figure}[t]
\centering
\includegraphics[height=10cm,width=0.9\textwidth,clip]{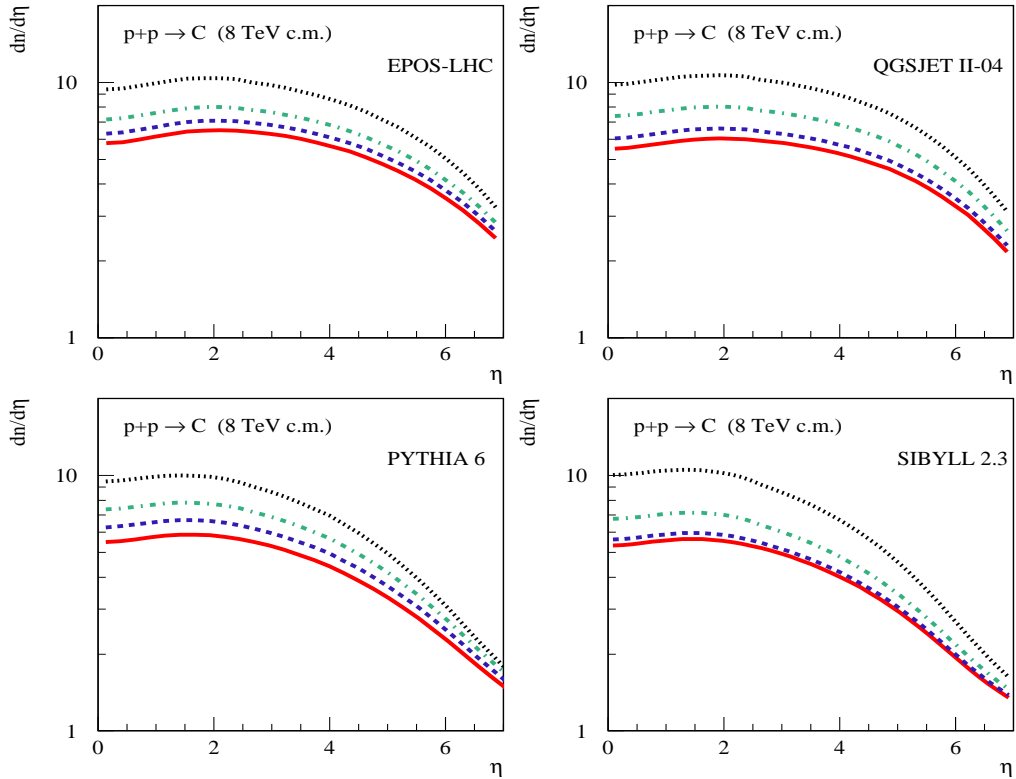}
\caption{Pseudorapidity   density of produced charged hadrons
 in $pp$ collisions at $\sqrt{s}=8$ TeV for different event selections:
 at least 1 (solid), 5 (dashed), 10 (dash-dotted), or  20 (dotted) charged
 hadrons of $p_{\rm t}>0.1$ GeV, produced  at $|\eta|<2.5$.
 Top left panel -- EPOS-LHC, top right panel --
 QGSJET-II-04, bottom left panel -- PYTHIA~6, bottom right panel --  
 SIBYLL~2.3.}
\label{fig:cms-tot8-diff}       
\end{figure}

\begin{figure}[t]
\centering
\includegraphics[height=10cm,width=0.9\textwidth,clip]{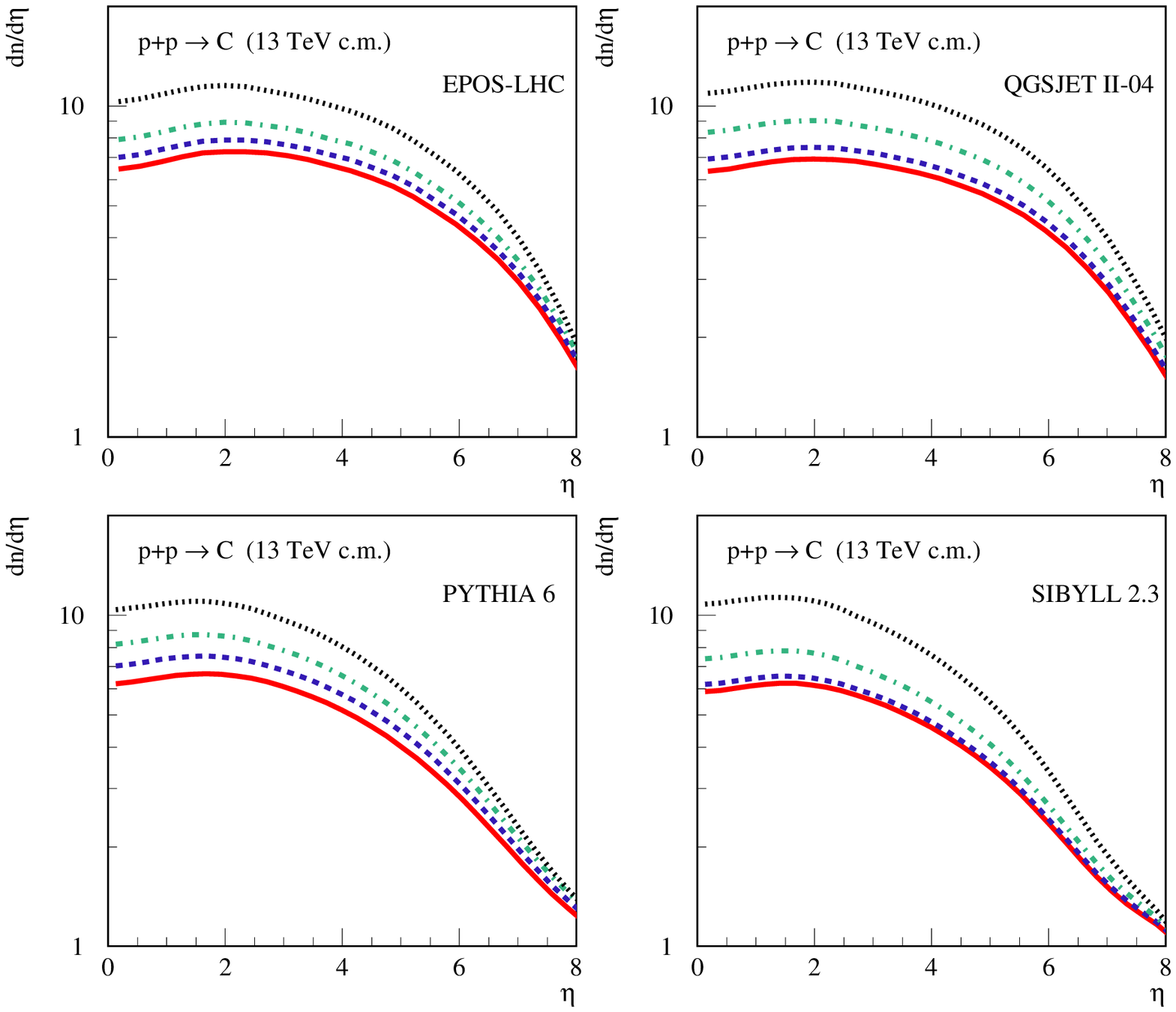}
\caption{Same as in  Fig.\  \ref{fig:cms-tot8-diff} for  $pp$ collisions at 
 $\sqrt{s}=13$ TeV.}
\label{fig:cms-tot13-diff}       
\end{figure}

As follows from the discussion in Section~\ref{intro.sec}, a better way 
  to discriminate between the two approaches for the treatment of 
  constituent parton Fock states  is via combined measurements of particle
production by central and forward-looking detectors, triggering different
levels of hadronic activity in the former. This is illustrated in 
 Figs.\  \ref{fig:cms-tot8-diff} and \ref{fig:cms-tot13-diff} for $pp$ collisions %
at $\sqrt{s}=8$  and 13 TeV, where the predictions of the different models for
$dn^{\rm ch}_{pp}/d\eta$ are compared to each other for different event 
 triggers: requesting at least 1, 5, 10, or 20 charged hadrons 
 of $p_{\rm t}>0.1$ GeV to be  detected at $|\eta|<2.5$. While all the considered
 models agree approximately with each other in the central rapidity region,
 the results of the models diverge considerably at large $\eta$.
 In PYTHIA~6 and SIBYLL~2.3, enhancing the trigger conditions gives rise to the
  increasing density of produced hadrons at small $\eta$  while leaving
  $dn^{\rm ch}_{pp}/d\eta$ practically unchanged in the large $\eta$ range.
  This is related to the fact that
 multiple scattering affects mostly the central particle
 production in the two models,
 while having a weak influence on forward hadron spectra.
 In contrast, in EPOS-LHC and  QGSJET-II-04 also the forward $\eta$-density
 of produced hadrons is substantially enhanced, reflecting the strong
 correlations between central and forward hadron production in the two models:
 A stronger central activity implies an increasing role of proton
  Fock states containing numerous large $x$ constituent partons.
 
\begin{figure}[t]
\centering
\includegraphics[height=5.5cm,width=0.9\textwidth,clip]{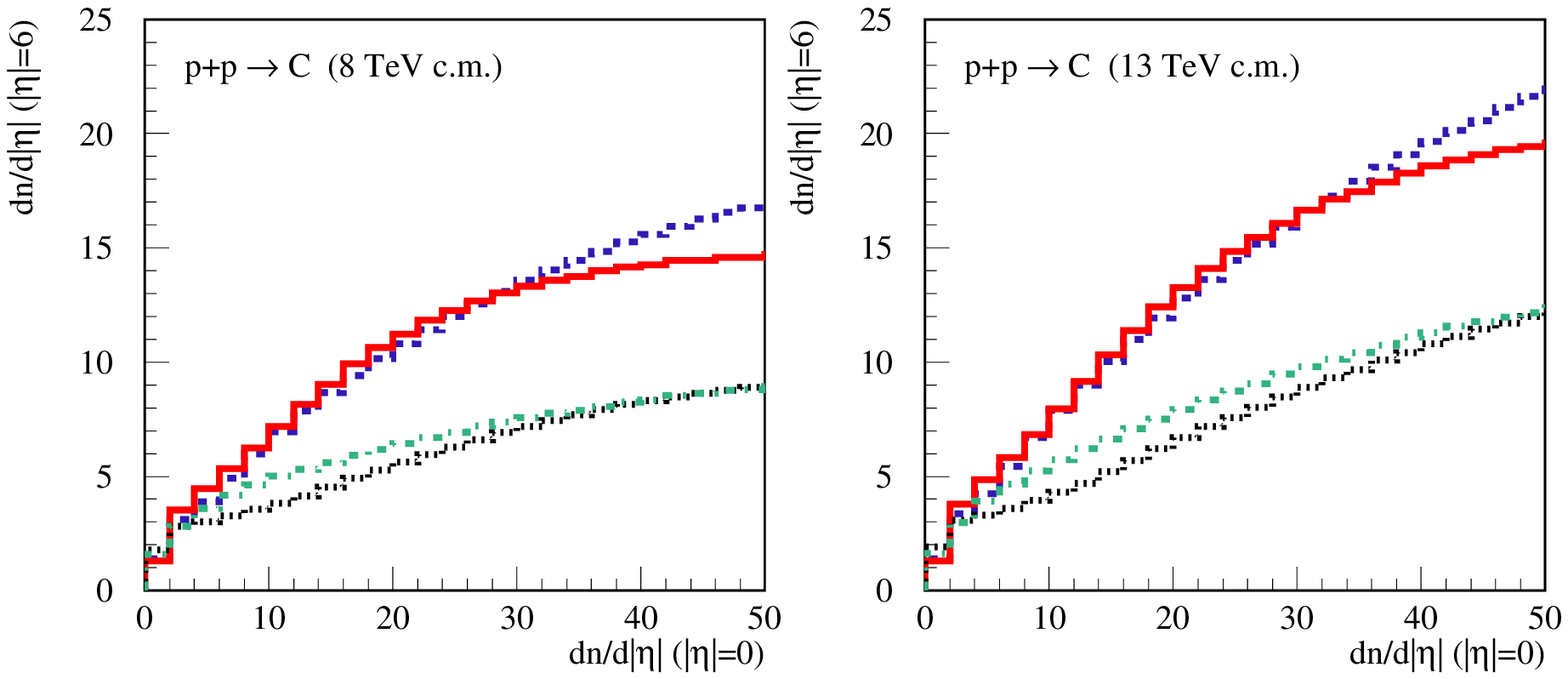}
\caption{Pseudorapidity   density of produced charged hadrons
$dn^{\rm ch}_{pp}/d|\eta|$ at  $|\eta|=6$ ($p_{\rm t}>0$) 
as a function of $dn^{\rm ch}_{pp}/d|\eta|$ at  $|\eta|=0$
 ($p_{\rm t}>0.1$ GeV)
  in $pp$ collisions at $\sqrt{s}=8$ TeV (left) and  $\sqrt{s}=13$ TeV (right),
  as calculated using EPOS-LHC (solid),
 QGSJET-II-04 (dashed), PYTHIA~6 (dash-dotted), and SIBYLL~2.3 (dotted).}
\label{fig:cms-tot-corr}       
\end{figure}

 The differences between the two basic approaches are best seen in the 
 correlations of the signal strength in central and forward-looking detectors,
 which may be studied by the CMS and TOTEM experiments. This is illustrated
 in  Fig.\  \ref{fig:cms-tot-corr}, 
 where we plot for  $\sqrt{s}=8$ and 13 TeV the $\eta$-density of
 charged hadrons $dn^{\rm ch}_{pp}/d|\eta|$
 (for arbitrary transverse momenta) at $|\eta|=6$
  (averaged over the interval $5.5<|\eta|<6.5$) 
  as a function of the central pseudorapidity density of
  charged hadrons of $p_{\rm t}>0.1$ GeV
 (averaged over the interval $|\eta|<1$). Both EPOS-LHC and QGSJET-II-04 predict
a strong correlation of the signal strength in CMS and TOTEM. The respective
results of the two models practically coincide with each other, apart from the
region corresponding to the tails of the multiplicity distributions.
In contrast, for PYTHIA~6  and SIBYLL~2.3 such a correlation turns out to be
twice weaker, being thus a clear experimental signature for discrimination
between the two theoretical approaches depicted schematically in
Fig.~\ref{fig:fock}.

\begin{figure}[t]
\centering
\includegraphics[height=10cm,width=0.9\textwidth,clip]{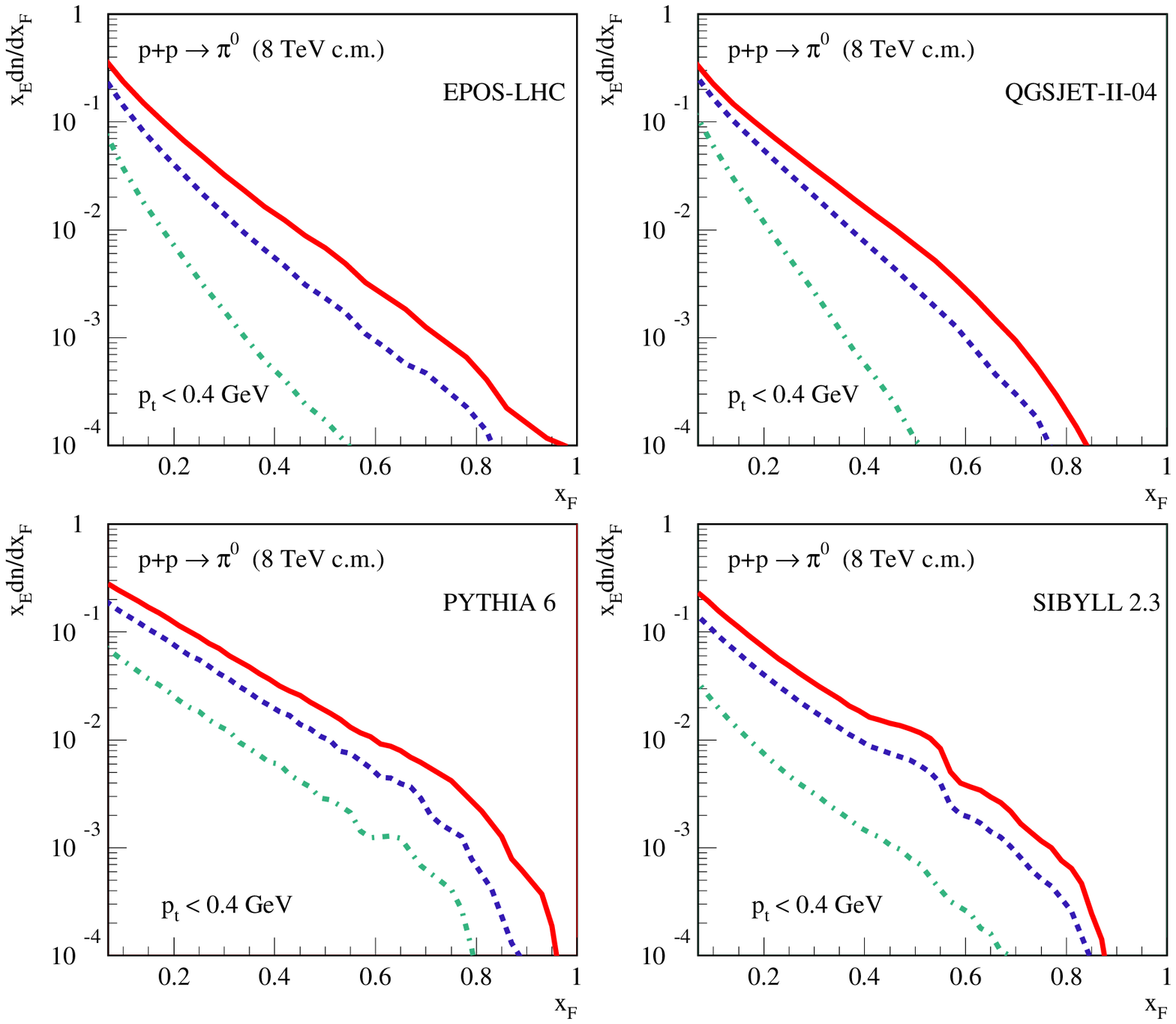}
\caption{Feynman $x$ spectra of neutral pions ($p_{\rm t}<0.4$ GeV)
 in $pp$ collisions at $\sqrt{s}=8$ TeV for different event selections:
 at least 1 (solid), 6 (dashed), or 20 (dash-dotted) charged
 hadrons of $p_{\rm t}>0.5$ GeV, produced  at $|\eta|<2.5$.
 Top left panel -- EPOS-LHC, top right panel --
 QGSJET-II-04, bottom left panel -- PYTHIA~6, bottom right panel --  
 SIBYLL~2.3.}
\label{fig:lhcf-pi8}       
\end{figure}

\begin{figure}[t]
\centering
\includegraphics[height=10cm,width=0.9\textwidth,clip]{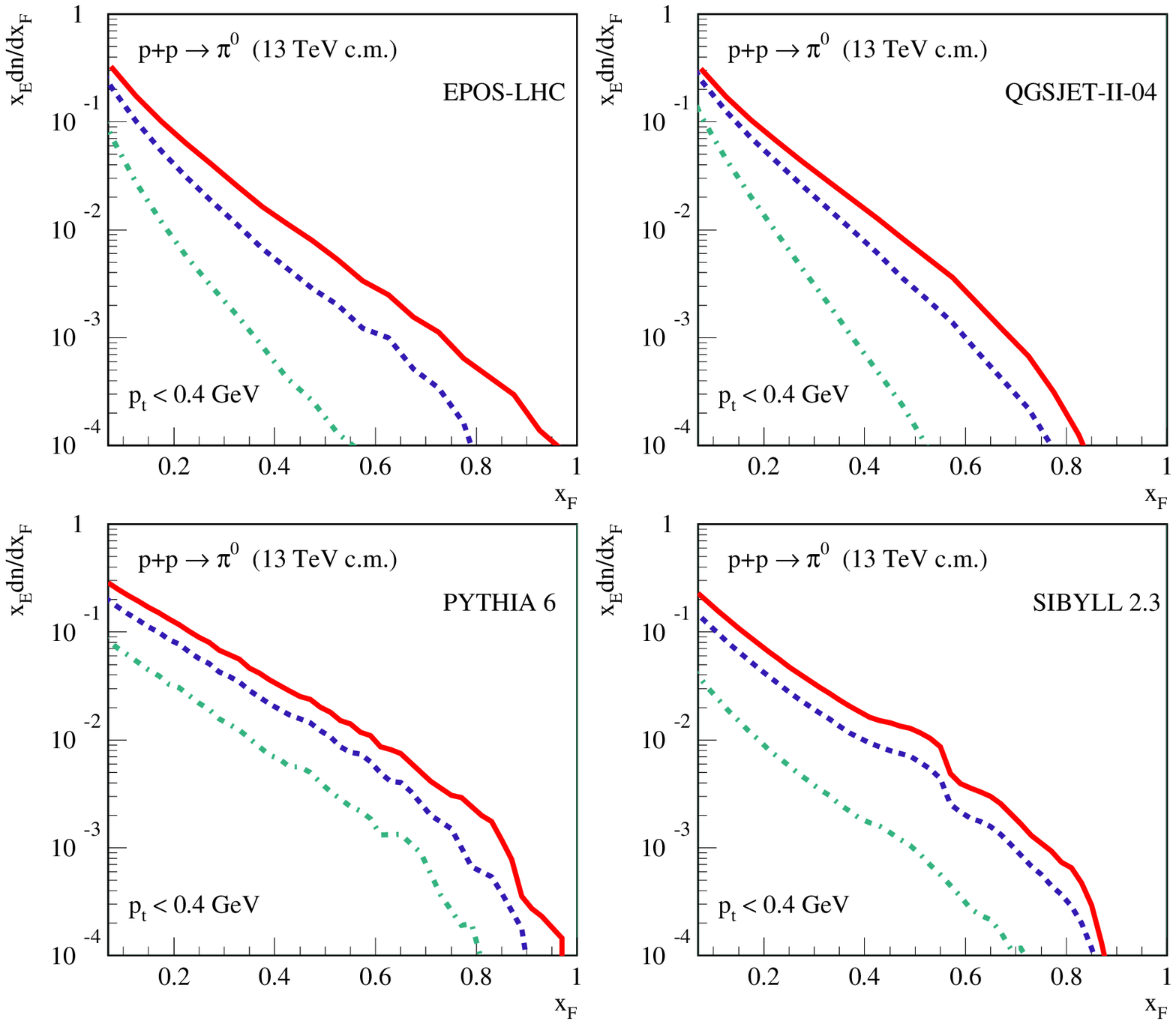}
\caption{Same as in  Fig.\  \ref{fig:lhcf-pi8} for  $pp$ collisions at 
 $\sqrt{s}=13$ TeV.}
\label{fig:lhcf-pi13}       
\end{figure}

Another way for the model discrimination may be provided by measurements
 of very
forward particle production, e.g.\ by the LHCf experiment, when supplemented
by triggering different hadronic activities in the central detectors of ATLAS.
This is illustrated in  Figs.\  \ref{fig:lhcf-pi8} and \ref{fig:lhcf-pi13}, where %
forward spectra of neutral pions in $pp$ collisions at $\sqrt{s}=8$ and 13 TeV
are shown for a number of ATLAS triggers, 
calculated using the different interaction models. It is easy to see that
SIBYLL and PYTHIA predict  an almost perfect limiting fragmentation:
For all the triggers considered, the spectral shapes remain practically
identical at large Feynman $x$, the spectra being just rescaled downwards
according to the corresponding event rates. This is a direct consequence
of the decoupling of central and forward production in the two models: 
triggering
more hadronic activity in the central detectors of
ATLAS enhances the central production while having
almost no impact on the forward hadron spectra. The picture changes
drastically in the case of the EPOS-LHC and  QGSJET-II-04 models characterized
by strong long-range correlations between central and forward hadron
production. Events with higher  multiplicity in ATLAS, corresponding to
enhanced multiple scattering, are dominated by contributions of Fock states
with larger numbers of large $x$ constituent  partons. The energy-momentum
sharing between these partons leaves its imprint on the very forward spectra
of $\pi^0$, which become noticeably softer, compared to the case of low
multiplicity events.

\begin{figure}[t]
\centering
\includegraphics[height=10cm,width=0.9\textwidth,clip]{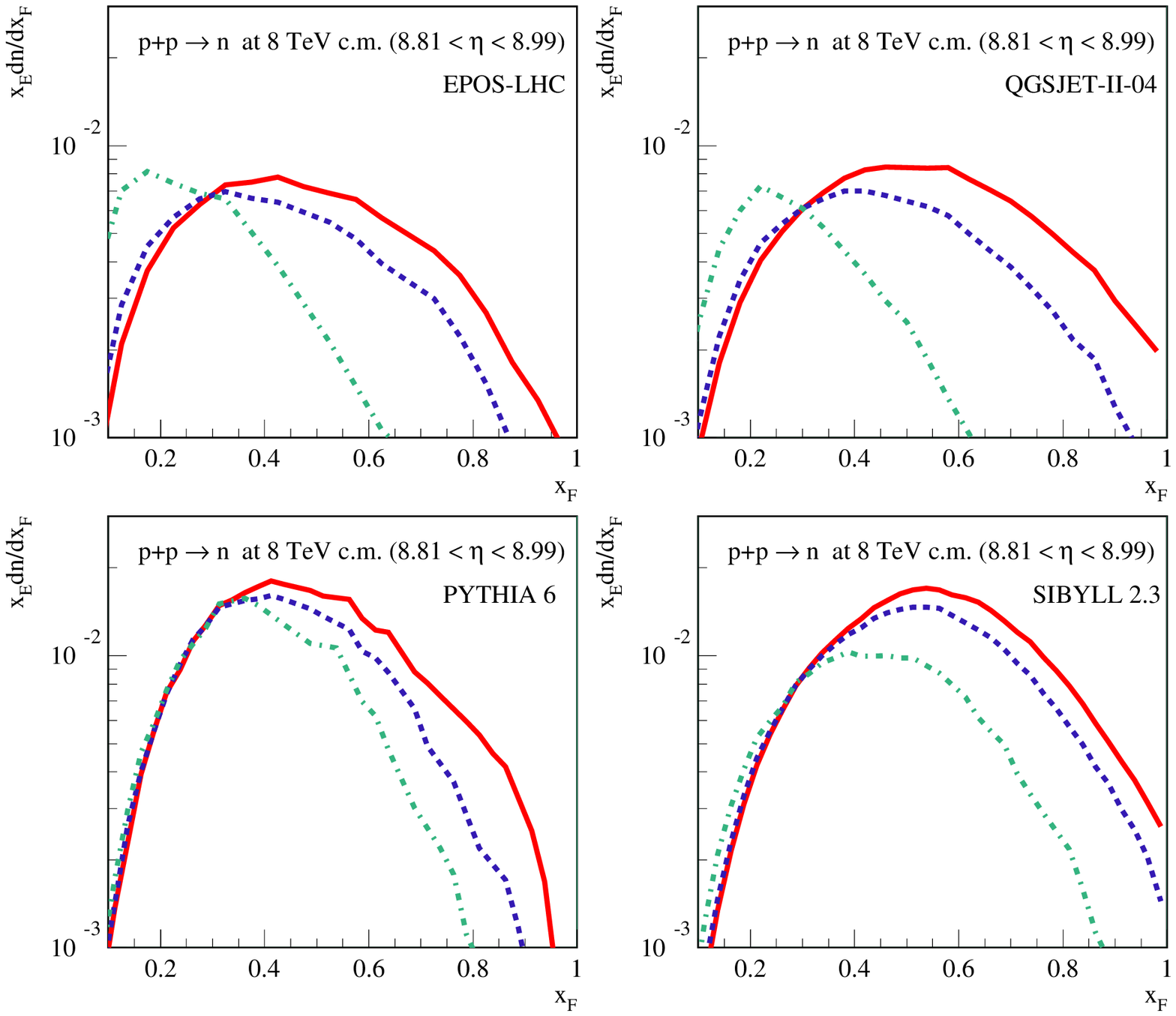}
\caption{Feynman $x$ spectra of neutrons ($8.81<\eta<8.99$)
 in $pp$ collisions at $\sqrt{s}=8$ TeV for different event selections:
 at least 1 (solid), 6 (dashed), or 20 (dash-dotted) charged
 hadrons of $p_{\rm t}>0.5$ GeV, produced  at $|\eta|<2.5$.
 The spectra for the latter two selections, plotted by dashed and dash-dotted
 lines, are rescaled to intersect the solid lines at $x_{\rm F}=0.3$.
 Top left panel -- EPOS-LHC, top right panel --
 QGSJET-II-04, bottom left panel -- PYTHIA~6, bottom right panel --  
 SIBYLL~2.3.}
\label{fig:lhcf-n8}       
\end{figure}

\begin{figure}[t]
\centering
\includegraphics[height=10cm,width=0.9\textwidth,clip]{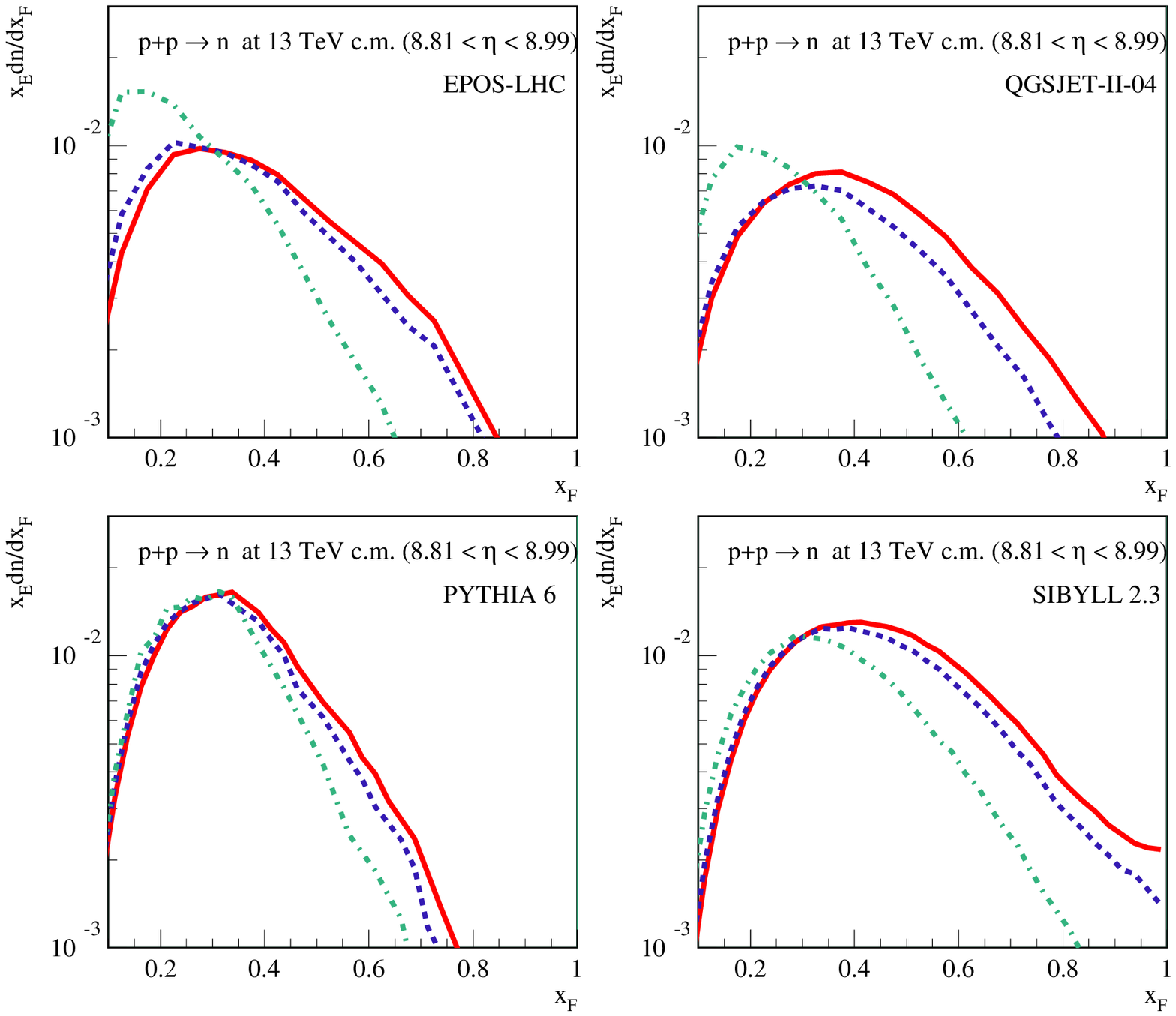}
\caption{Same as in  Fig.\  \ref{fig:lhcf-n8} for  $pp$ collisions at 
 $\sqrt{s}=13$ TeV.}
\label{fig:lhcf-n13}       
\end{figure}

An even more drastic consequence of the energy-momentum
sharing between large $x$  constituent   partons is the considerable
softening of forward nucleon spectra. Higher rates of multiple scattering,
corresponding to enhanced hadronic activity in ATLAS, require larger numbers
of  constituent   partons to be involved in particle production. Hence, 
smaller fractions of the initial proton momentum are left for spectator partons
which finally form the ``leading'' (most energetic) nucleons. This effect is
clearly seen  in the predictions of  EPOS-LHC and  QGSJET-II-04 for the
forward spectra of neutrons in $pp$ collisions at $\sqrt{s}=8$ and 13 TeV, as
  shown in Figs.\  \ref{fig:lhcf-n8} and \ref{fig:lhcf-n13} %
for the pseudorapidity range covered by the LHCf experiment and for
the same selection   of ATLAS triggers as above. On the contrary,
  the shapes of  forward neutron spectra predicted by
SIBYLL and PYTHIA demonstrate much weaker sensitivity to varying the trigger
conditions, as one may see in   Figs.\  \ref{fig:lhcf-n8} and \ref{fig:lhcf-n13}.
This is because the additional constituent partons involved
in multiple scattering are   the low $x$ ones. Hence, they typically
pick up small fractions of the initial proton momenta, thus having a weak
impact on the formation of leading nucleons.

\section{Discussion}
\label{sec:Discussion}

\begin{figure}[htb]
\centering
\includegraphics[height=5.5cm,width=0.45\textwidth,clip]{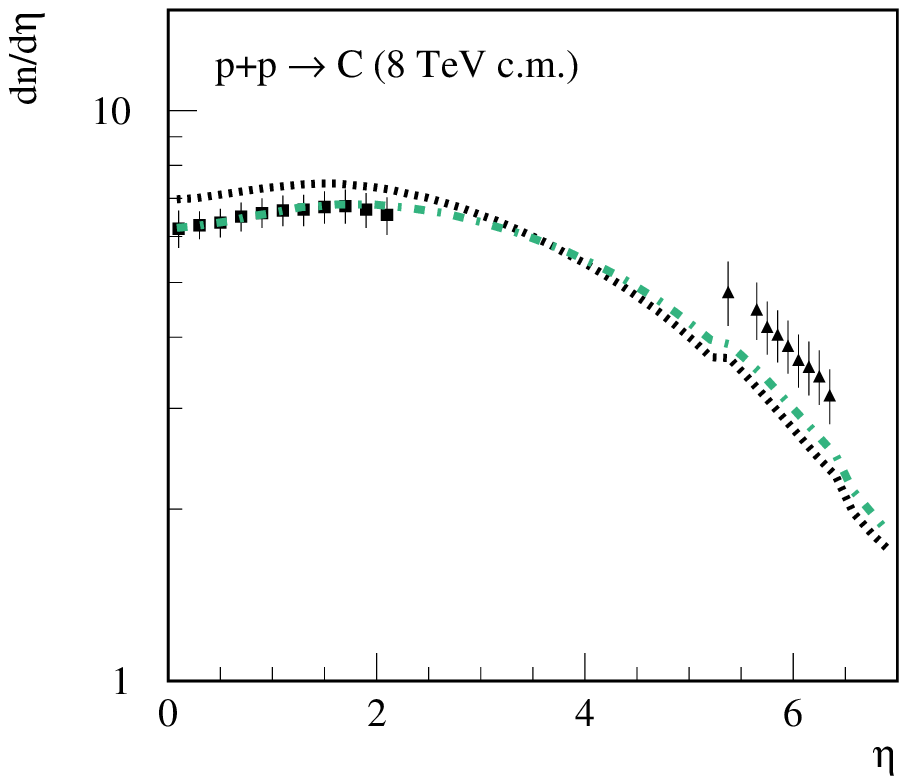}
\caption{Same as in Fig.\ \ref{fig:cms-tot} (left) for the SIBYLL~2.3 (dotted)
and SIBYLL~2.1 (dash-dotted) models.}
\label{fig:cms-sib}       
\end{figure}

\begin{figure}[htb]
\centering
\includegraphics[height=5.5cm,width=0.45\textwidth,clip]{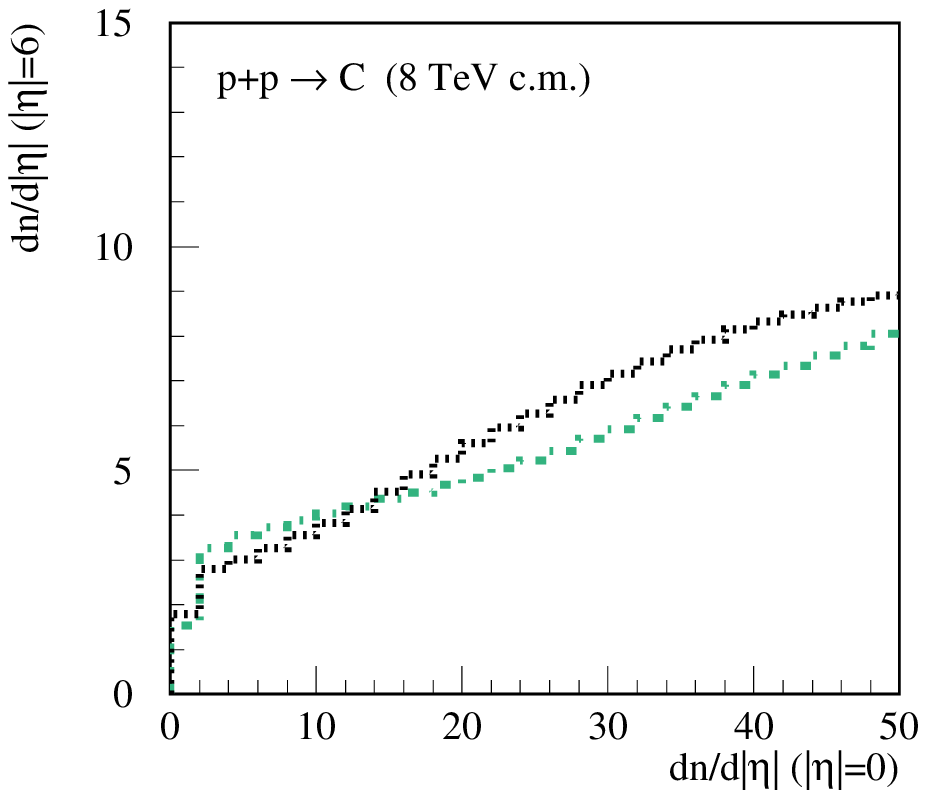}
\caption{Same as in Fig.\ \ref{fig:cms-tot-corr} (left) for the 
SIBYLL~2.3 (dotted) and SIBYLL~2.1 (dash-dotted) models.}
\label{fig:cms-sib-corr}       
\end{figure}

\begin{figure}[htb]
\centering
\includegraphics[height=5.5cm,width=0.45\textwidth,clip]{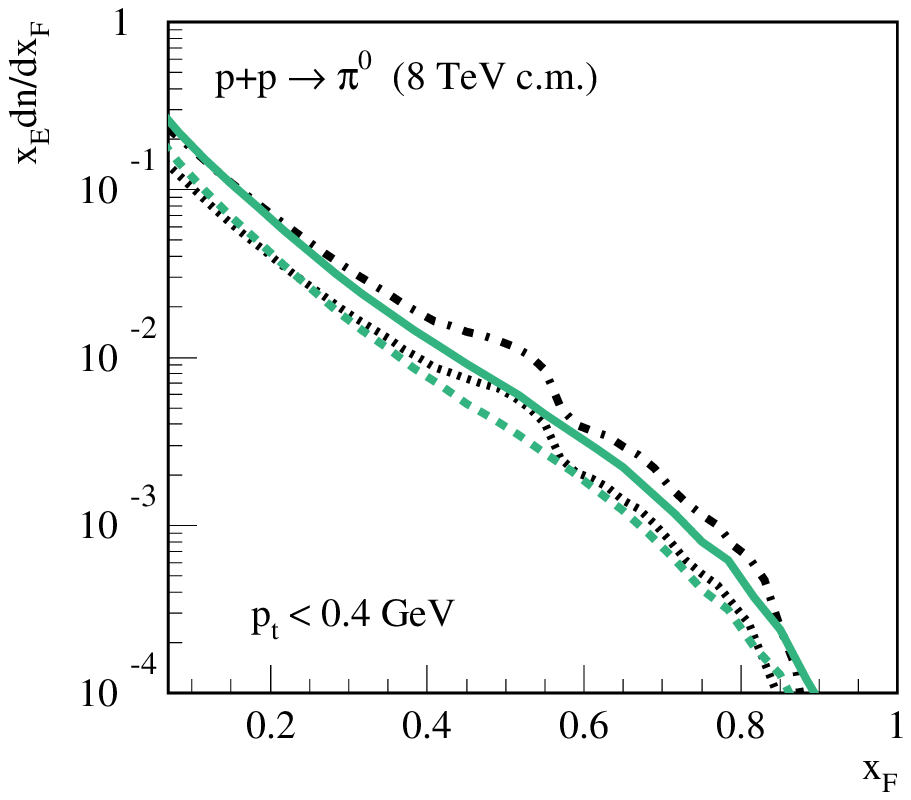}
\caption{Feynman $x$ spectra of neutral pions ($p_{\rm t}<0.4$ GeV)
 in $pp$ collisions at $\sqrt{s}=8$ TeV, as calculated using the
SIBYLL~2.3   and SIBYLL~2.1   models   for different event selections:
 at least 1 (solid -- SIBYLL~2.1, dash-dotted -- SIBYLL~2.3) or 6
  (dashed -- SIBYLL~2.1, dotted -- SIBYLL~2.3) charged
 hadrons of $p_{\rm t}>0.5$ GeV, produced  at $|\eta|<2.5$.}
\label{fig:lhcf-sib}       
\end{figure}

One may ask here  natural questions about whether the above-discussed 
observables
 are robust enough with respect to a retuning of   model parameters and
 how sensitive   they are to other features of the interaction treatment.
 We may get some feeling about this by comparing in 
 Figs.~\ref{fig:cms-sib}--\ref{fig:lhcf-sib}  
 the results of
   SIBYLL~2.3 and   the pre-LHC version of the model
    (SIBYLL~2.1 \cite{ahn09}) for
  $pp$ collisions at $\sqrt{s}=8$ TeV   for the ND $\eta$-density of charged particles, for the
correlation of $dn^{\rm ch}_{pp}/d|\eta|$ at  $|\eta|=6$ and
  $\eta=0$, and for forward spectra of neutral pions as a function
  of particle activity in the central detector of ATLAS.
  As we can see from the figures,  both the $\eta$-dependence
  in Fig.~\ref{fig:cms-sib} and the $x_{\rm F}$ spectral shapes for 
  particular ATLAS triggers
  in Fig.~\ref{fig:lhcf-sib} demonstrate a noticeable dependence on the tuning.
  Nevertheless, both model versions predict a similarly weak correlation
   between   the $\eta$-densities of produced charged hadrons at small and 
   large $\eta$, as shown   in Fig.~\ref{fig:cms-sib-corr}.
  Similarly, both tunes demonstrate
  an almost perfect limiting fragmentation: The shape of the forward 
   $x_{\rm F}$-spectrum  of produced neutral pions in Fig.~\ref{fig:lhcf-sib}
    remains remarkably   insensitive
to the triggered particle activity in the ATLAS detector.

Similar conclusions follow from comparing the results of EPOS-LHC and QGSJET-II-04 with each other. Due to the large differences between the
model approaches concerning  the treatment of nonlinear interaction
effects and due to the collective effects implemented in the hadronization
procedure of the EPOS model, their predictions both for the inelasticity
$K^{\rm inel}_{pp}$  in Fig.\ \ref{fig:inel} and for the forward spectra
of $\pi^0$s and neutrons in Figs.\ \ref{fig:lhcf-pi8}--\ref {fig:lhcf-n13}
differ considerably. Nevertheless, both models predict a   similarly strong
correlation between the strengths of particle production in the central and
fragmentation regions (see Fig.\ \ref{fig:cms-tot-corr}) and a strong violation
of the limiting fragmentation for very forward hadron production -- both 
being the consequences of the similar model assumptions concerning the structure
of constituent parton Fock states,   discussed in Section \ref{sec:intro}.

Let us finally discuss the question of whether
 the differences between, e.g., PYTHIA~6
and SIBYLL~2.3 for the proposed observables give us the characteristic
scale for the Monte Carlo tuning dependence of the presented results.
In a sense, this is indeed the case -- 
as these models represent the two extreme
approaches corresponding to the mechanism of    Fig.\ \ref{fig:fock} (left).
Indeed, in the SIBYLL model all multiple scattering processes except the
first one are treated as binary gluon-gluon collisions, the initial state
 radiation (ISR) of partons being neglected. Hence, the decoupling between
  the central and forward particle production is maximal in that model.
   On the other hand, ISR in PYTHIA  is traced backwards {\it for each
hard scattering process independently}, down to the transverse momentum
cutoff $p_{\rm t}^{\rm cut}\sim$~few GeV. At first sight, one would tend
to classify that model as corresponding to the picture in  Fig.\
 \ref{fig:fock} (right). However, because the ISR in PYTHIA is matched to the
 parton distribution functions (PDFs) of gluons and sea quarks at the relatively high  $p_{\rm t}^{\rm cut}$ scale, the model corresponds
  effectively to the mechanism of    Fig.\ \ref{fig:fock} (left): Due to the
  steep low $x$ rise of the respective PDFs and their strong suppression at
  $x\rightarrow 1$, the light-cone momenta of  end-point partons in the backward cascades are sampled predominantly as $\propto 1/x$.
  
   In relation  to the latter,
   it is worth mentioning the recently developed Monash tune of the
  PYTHIA~8 generator \cite{ska14}, whose results agree well with the 
  above-discussed data  of  CMS and TOTEM on  $dn^{\rm ch}_{pp}/d\eta$
  and are generally similar to the ones of EPOS-LHC for forward hadron
  production in the kinematic range studied by the CMS experiment
  \cite{cms-forward}. Among other improvements, the Monash tune
  employs the new NNPDF2.3LO PDF set \cite{bal13} characterized by a rather
  hard gluon PDF. We believe it is the latter choice which governs the new
  model predictions for forward hadron production: Due to the presence of the
  ``valencelike'' contribution in the gluon PDF, 
  initial state parton cascades for different
  multiple scattering processes end up with  somewhat  harder 
  (i.e.\ having  larger momentum fraction $x$) gluons and the model
 moves closer to the picture in  Fig.\
 \ref{fig:fock} (right).\footnote{Of importance here is 
 not only that the hard gluon PDF is chosen but that it is applied to sample
 light-cone momenta for  end-point gluons
 {\it in all the multiple initial state parton cascades}.}
  Thus, a comparison of results of different tunes of 
 PYTHIA~8, using different PDF sets, for the above-proposed observables
 is of interest.

\section{Conclusions}
\label{sec:conclusions}
One of the most crucial differences between current Monte Carlo generators of
high energy hadronic collisions is related to the
underlying assumptions concerning the structure of
constituent parton Fock states in hadrons.
These dominate the differences in model predictions concerning the
 energy dependence and target mass dependence of forward hadron production
 in hadron-proton and hadron-nucleus collisions. In  close relation to that,
 these model differences  have a crucial impact on the present theoretical
 uncertainties for basic characteristics of cosmic ray-induced extensive
 air showers, and hence also on the prospects of further success of studies
 of the primary composition of ultrahigh energy cosmic rays.
We demonstrated that the different model approaches can be discriminated
by measuring long-range correlations between central and forward hadron
production. In particular, our analysis shows that combined studies of 
proton-proton collisions at the LHC by central and forward-looking particle
detectors  have a rich potential for such a discrimination. 
It is worth stressing that such studies do not generally require
new measurements at the LHC -- as the corresponding analysis can be performed
using the already recorded experimental information from the common
data-taking by  CMS and TOTEM or, respectively, by   ATLAS and LHCf.

\subsection*{Acknowledgments}
This work was supported in part by
 Deutsche Forschungsgemeinschaft (Grant No.\
OS 481/1) and  the State of Hesse via the LOEWE-Center HIC for FAIR.

 \end{document}